\documentstyle[aps,twocolumn,epsfig]{revtex}
\newcommand{\beq}{\begin{equation}}
\newcommand{\eeq}{\end{equation}}
\newcommand{\bea}{\begin{eqnarray}}
\newcommand{\eea}{\end{eqnarray}}
\newcommand{\ba}{\begin{array}}
\newcommand{\ea}{\end{array}}
\newcommand{\bc}{\begin{center}}
\newcommand{\ec}{\end{center}}

\newcommand{\gsimeq}{\stackrel{>}{\scriptstyle\sim}}

\newcommand{\bml}{\begin{mathletters}}
\newcommand{\eml}{\end{mathletters}}
\newcommand{\commentout}[1]{{}}

\newcommand{\p}{{\bf p}}
\newcommand{\r}{{\bf r}}

\newcommand{\half}{\hbox{$1\over2$}}

\newcommand{\rb}{$^{85}$Rb }
\newcommand{\etal}{{\it et al.}}
\newcommand{\eq}[1]{(\ref{#1})}

\newcommand{\phidag}{\phi^\dagger}
\newcommand{\psidag}{\psi^\dagger}

\begin{document}
\wideabs
{
\title{Mean-Field Theory of Feshbach-Resonant
Interactions in \rb Condensates}

\author{Matt Mackie,$^1$ Kalle-Antti
Suominen,$^{1,2}$ and Juha Javanainen$^3$}
\address{$^1$Helsinki Institute of Physics, PL 64,
FIN-00014
Helsingin yliopisto, Finland \\
$^2$Department of Physics, University of
Turku, FIN-20014 Turun yliopisto, Finland \\
$^3$Physics Department, University of Connecticut,
Storrs,
Connecticut, 06269-3046, USA}

\date{\today}

\maketitle

\begin{abstract}
Recent Feshbach-resonance experiments with \rb
Bose-Einstein
condensates have led to a host of unexplained results:
dramatic
losses of condensate atoms for an across-resonance
sweep of the
magnetic field, a collapsing condensate with a burst
of
atoms emanating from the remnant condensate, increased
losses
for decreasing interaction times---until very short
times are reached, and
coherent oscillations between remnant and burst
atoms. Using a
simple yet realistic mean-field model, we find that
rogue
dissociation, molecular dissociation to noncondensate
atom pairs, is strongly implicated as the physical
mechanism responsible
for these observations.
\end{abstract}
\pacs{PACS numbers: 03.75.Fi,03.65.Bz}
}

{\it Introduction}-- The process known as the Feshbach
resonance occurs when two ultracold atoms collide in
the
presence of a magnetic field, whereby a spin flip of
one atom
can induce the pair to jump from the two-atom
continuum to a
quasibound molecular state. If the initial atoms are
Bose condensed, the so-formed molecules
will also
comprise a Bose-Einstein
condensate~(BEC)~\cite{MBEC_FR}. Since the
Feshbach resonance is mathematically identical to
photoassociation~\cite{DRU98,JAV99,KOS00}, the process
that occurs when
two ultracold atoms form a molecule by absorbing a
photon, insight
gathered in either case is applicable to the other. In
particular, the most recent results from
photoassociation predict that
rogue dissociation, molecular dissociation to
noncondensate atom pairs,
imposes a maximum achievable rate on atom-molecule
conversion, as well as
the possibility of coherent Rabi oscillations between
the BEC
and the dissociated atom pairs~\cite{JAV02}.

As predicted~\cite{TUNE}, the Feshbach resonance was
used to enable Bose condensation in
\rb by
tuning the natively negative atom-atom scattering
length into
the positive regime~\cite{COR00}. Part of this experiment
involved a sweep of
the magnetic field across
the Feshbach resonance, resulting in heavy
condensate losses ($\sim80$\% for the slowest sweep
rates).
Additional experiments led to the observation of
collapsing
condensates, an event characterized by bursts of atoms
emanating from a remnant BEC, and coined
''bosenova" for the
analogy with a supernova explosion~\cite{DON01}. More recently,
experiments with pulsed magnetic fields that come
close to, but
do not cross, the Feshbach-resonance have revealed a
striking increase in condensate loss for a decrease in
the
interaction time-- until reaching very short times~\cite{CLA02}. Finally,
double-pulse results
indicate large amplitude remnant-burst oscillations
($\sim 25\%$), with
the missing atoms prompting speculation on the
formation of a
molecular condensate~\cite{DON02}.

However, recent work on Feshbach-stimulated
photoproduction of stable molecular condensates
indicates that rogue dissociation dominates
atom-molecule
conversion for the above \rb Feshbach
resonance, meaning that it is practically useless for
producing any
significant fraction of molecular BEC, stable or otherwise~\cite{MAC02}.
Additionally, assertions of a breakdown of
mean-field theory in the face of
large resonance-induced scattering
lengths~\cite{COR00,CLA02} are intriguing, especially given that theory
actually faults the effective all-atom
description~\cite{MFT_VALID}. The purpose of this
Letter is to test a
mean-field model of coherent atom-molecule conversion
against the salient
features of the JILA experiments, with the intent of
developing an understanding of the basic physics
involved.

{\it Mean-field theory}-- The mathematical equivalence
of the
Feshbach-resonance~\cite{MBEC_FR} and
photoassociation~\cite{DRU98,JAV99}
lies in both processes being described, in
terms of second
quantization, as destroying two atoms and creating a
molecule. We
therefore model a quantum degenerate gas of atoms
coupled via a Feshbach
resonance to a condensate of quasibound molecules
based on
Refs.~\cite{JAV99,KOS00,JAV02}. The initial atoms are
denoted by the
boson field
$\phi(\r,t)$, and the quasibound molecules by the
field
$\psi(\r,t)$. The Hamiltonian density for this system
is
\bea {{\cal H}\over\hbar} &=&
\phidag\left[-{\hbar\nabla^2\over2m}\right]\phi\nonumber
+\psidag\left[-{\hbar\nabla^2\over4m}+\delta_0\right]\psi
\\&& -{\Omega\over2\sqrt{\rho}}\,[\psi^\dagger\phi\phi
+\phi^\dagger\phi^\dagger\psi] \,,
\label{CURHAM}
\eea
\beq
\Omega =
\lim_{\epsilon\rightarrow0}\sqrt{{\sqrt2\pi\hbar^{3/2}\rho\over
\mu^{3/2}}{\Gamma(\epsilon)\over\sqrt{\epsilon}}}\,,
\label{OMDEF}
\eeq
where $m=2\mu$ is the mass of an atom, $\hbar\delta_0$
is the energy
difference between a
molecule and two atoms, $\rho$ is an invariant
density equal to the sum of atom density and twice the
molecular
density, and
$\Gamma(\epsilon)$ is the dissociation rate for a
molecule with
the energy
$\hbar\epsilon$ above the threshold of the
Feshbach resonance.

Switching to momentum
space, only zero-momentum atomic and molecular
condensate modes are
retained, represented by the respective $c$-number
amplitudes
$\alpha$ and
$\beta$. We also take into account correlated pairs of
noncondensate
atoms using a complex amplitude
$C(\epsilon)$, which represent pairs of noncondensate
atoms in the manner
of the Heisenberg picture expectation value $\langle
a_\p
a_{-\p}\rangle$, with $\hbar\epsilon$ being the
relative energy of the
atoms.  The normalization of our mean fields is such
that
$|\alpha|^2+|\beta|^2+\int d\epsilon\,|C(\epsilon)|^2
= 1$.  We work from
the Heisenberg equation of motion of the boson
operators under the
simplifying assumption that the noncondensate atoms
pairs are only
allowed to couple back to the molecular condensate,
ignoring the
possibility that noncondensate atoms associate to make
noncondensate molecules. This neglect
is justified to the extent that
Bose enhancement favors transitions back to the
molecular condensate. The
final mean-field equations are~\cite{JAV02}
\bml
\bea
i\dot\alpha &=&
     -{\Omega\over\sqrt{2}}\alpha^*\beta, \\
i\dot\beta &=& \delta_0\beta
-{\Omega\over\sqrt{2}}\alpha\alpha
                  -{\xi\over\sqrt{2\pi}}\int
d\epsilon\,\sqrt[4]{\epsilon}\,C(\epsilon),
\\
i\dot C(\epsilon) &=& \epsilon C(\epsilon)
-{\xi\over\sqrt{2\pi}}\,\sqrt[4]{\epsilon}\,\beta\,.
\eea
\label{EQM}
\eml
The analog of the Rabi frequency for the rogue
modes $\xi$ is inferred using Fermi Golden rule, which
gives the dissociation rate for a positive-energy
molecule as
\beq
\Gamma(\epsilon)=\sqrt{\epsilon}\,\xi^2\,.
\label{GAMDEF}
\eeq

Next the problem is reformulated in terms of two key
parameters
with the dimension of frequency. The density-dependent
frequency
\beq
\omega_\rho = {\hbar\rho^{2/3}\over m},
\label{OMRHODEF}
\eeq
has been identified before, along with the
operational significance that, once
$\Omega\gsimeq\omega_\rho$, rogue dissociation is
expected to be a dominant factor in the
dynamics~\cite{JAV99,KOS00,JAV02}. Here it is
convenient to define another primary parameter with
the dimension of frequency. Considering on-shell
dissociation of molecules to atoms with
the relative energy
$\epsilon$, the Wigner threshold law delivers a
dissociation rate
$\Gamma(\epsilon)$ such that
$\Gamma(\epsilon)/\sqrt\epsilon$ converges to a finite
limit for
$\epsilon\rightarrow0$; hence, we define
\beq
\Xi = \left(\lim_{\epsilon\rightarrow 0}\,
{\Gamma(\epsilon)\over
2\sqrt\epsilon}
\right)^2,
\label{XIDEF}
\eeq
which indeed has the dimension of frequency.
Comparison of Eqs.~\eq{OMDEF}, \eq{GAMDEF},
\eq{OMRHODEF}
and~\eq{XIDEF} give the parameters in the mean-field
equations
as
\beq
\Omega = 2^{3/2}
\sqrt{\pi}\,\Xi^{1/4}\omega_\rho^{3/4}, \quad
\xi=\sqrt{2}\,\Xi^{1/4}\,.
\eeq

{\it Renormalization}-- When the coupling to the
continuum of
noncondensate atom pairs is included, the continuum
shifts the
molecular state~\cite{CONT_SHIFT}. As explained
before~\cite{JAV02}, and will be  discussed in more
detail
below, we have taken the dominant state pushing into
account in our calculations. However, there is a
relevant
residual effect to consider.

To begin with, consider the equation of motion for the
molecular
amplitude, including the coupling to noncondensate
atom pairs
but not to the atomic condensate [set $\Omega=0$ in
Eqs.~\eq{EQM}]. Again, the particular energy
dependence of
the
coupling comes from the Wigner threshold law, which is
here
assumed to be valid for all relevant energies; but, as
is necessary in the numerical calculations anyway, we
cut off
the coupling between molecules and atom pairs at some
frequency
$\epsilon_M$. The question of renormalization is the
question
of the dependence of the results on the
cutoff $\epsilon_M$.

Equations~\eq{EQM} (with $\Omega=0$) are easy to solve
using,
say, the Fourier transformation. With the initial
condition
that $\beta(t=0)=1$, for positive times the solution
has the
Fourier transform
\beq
\beta(\omega) =
{i\over\omega+i\eta-\delta_0-\Sigma(\omega)}\,,
\eeq
where the self-energy is
\beq
\Sigma(\omega) =
{\xi^2\over
{2\pi}}\int_0^{\epsilon_M}d\epsilon\,{\sqrt\epsilon
\over\omega+i\eta-\epsilon}\,,
\label{SIGMA}
\eeq
and $\eta=0^+$.
Now, a real pole of $\beta(\omega)$ corresponds to a
true
stationary state of the Hamiltonian. It turns out that
for
suitable detunings  there is a real pole $\omega\le0$,
which
obviously corresponds to the coupling-renormalized
energy of the
molecules.

Assuming that the value of $\omega$ is negative, the
integral
\eq{SIGMA} is carried out easily and the equation
giving the
pole becomes
\beq
\omega-\left(\delta_0-{\sqrt{\Xi\epsilon_M}\over\pi}\right)-{2\over\pi}
\sqrt{-\omega\Xi}\,
\arctan\left[
{\sqrt{\epsilon_M}\over\sqrt{-\omega}}\right]=0\,.
\eeq
The term involving the detuning is the main
contribution to the
renormalization. As we have done  before in our
numerical
calculations~\cite{JAV02},  we choose the bare
detuning
$\delta_0$ so that, for the given energy cutoff
$\epsilon_M$, the renormalized detuning
$\delta_0-\sqrt{\epsilon_M\Xi}/\pi$ attains the
desired value.
Hereafter we use the symbol $\delta$ for the
renormalized
detuning. This is the parameter that is varied by
changing the
laser frequency in photoassociation, or the magnetic
field in
the Feshbach resonance. We carry out the rest of the
renormalization by setting
$\epsilon_M\rightarrow\infty$. We find
the equation giving the characteristic frequency
corresponding
to the molecules as
$
\omega - \delta - \sqrt{-\omega\Xi} = 0\,.
$
The proper solution is
\beq
\omega = \delta
+\half\sqrt\Xi\left(\sqrt{\Xi-4\delta}-\sqrt\Xi
\right)\,,
\label{SIGEQ}
\eeq
which is valid for all negative detunings and gives
the
characteristic frequency of the molecule.
Nevertheless, complete
diagonalization of the problem (with $\Omega=0$) shows
that
the mode evolving at this frequency is not the
original
or ``bare'' molecules, but a coherent superposition of
molecules
and noncondensate atoms pairs.

{\it Numerical procedures}-- The mean-field
equations~\eq{EQM} are
integrated using the norm-preserving
predictor-corrector algorithm
described in Ref.~\cite{JAV02}. Magnetic fields are
converted to detunings according to
$\hbar\delta=\Delta\mu(B_0-B)$, where the position
of the Feshbach
resonance is $B_0=154.9\,{\rm G}$, and where the
difference in magnetic
moments between bound molecules and free atom pairs,
$\Delta_\mu\approx2\,\mu_B$
    (where $\mu_B$ is the Bohr magneton), is borrowed
from
${}^{87}$Rb~\cite{WYN00}. Compared
to the ensuing detunings $\delta$, the interactions
energies
between the atoms due to the background scattering
length
$a=23.8\,{\rm nm}$ are immaterial. We therefore ignore
atom-atom
interactions unrelated to the Feshbach resonance, as
well as the
(unknown) atom-molecule and molecule-molecule
interactions.

We have used a number of different methods to estimate
the
parameter $\Xi$, all giving similar values. The
present argument
goes as follows. One of the experiments in
Ref.~\cite{DON02} gives
the characteristic frequency (presumably) of
molecules,
$\omega=
-2.07\times10^5\,{\rm s}^{-1}$, at the magnetic field
$B=159.69\,{\rm G}$ corresponding to the detuning
$\delta=-8.42\times10^7\,{\rm s}^{-1}$. Solving from
Eq.~\eq{SIGEQ}, we have $\Xi=5.29\times10^9\,{\rm
s}^{-1}$, and
thus $\xi=381\,{\rm s}^{-1/4}$. It should be
noted that, while our rogue-dissociation coupling can
be shown to give
the correct atom-atom scattering theory close to the
resonance
($\delta\rightarrow0$),  Eq.~\eq{SIGEQ}  will not
correctly reproduce the
molecular energy on the side of large magnetic fields
in Fig.~5 of
Ref.~\cite{DON02}. Passable agreement could be
reached by treating the
energy cutoff
$\epsilon_M$  as a  variable finite parameter;
unfortunately, the
abrupt cutoff at $\epsilon_M$ introduces physical and
numerical artifacts, which need to be cleaned up via a
judicious
energy dependence of the coupling. For example, the
entire
experimental range of magnetic fields can be fit by
multiplying the
original coupling by an exponential
[$\exp(-\epsilon/\epsilon_M)$] or
rational [${\epsilon_M^2/(\epsilon+\epsilon_M)^2}$]
cutoff. As our aims
are strictly qualitative, we discuss these details
elsewhere.

\begin{figure}
\centering
\epsfig{file=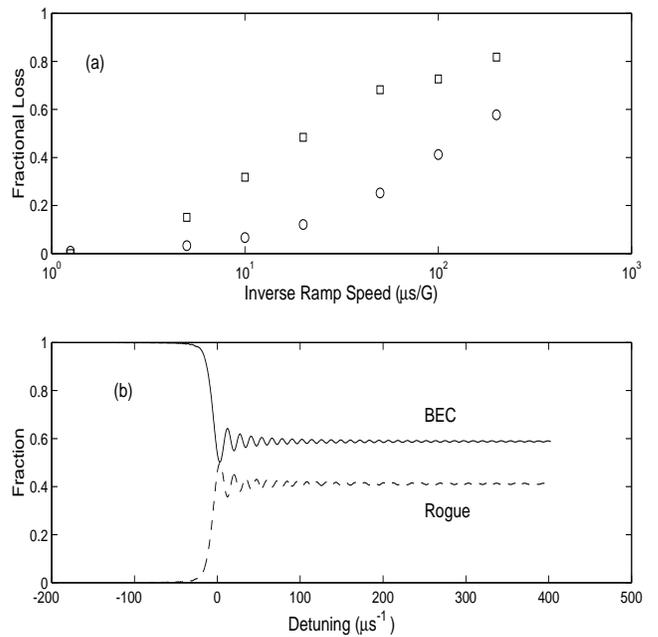,width=8.5cm,height=8.5cm}
\caption{(a) Experimental~\protect\cite{COR00}
($\Box$) and theoretical
($\circ$) atom loss incurred in sweeping a $^{85}$Rb
BEC across the
Feshbach resonance, where the magnetic field
is swept in a
linear fashion from $B_i=162$~G to $B_f=132$~G. In
each numerical run, the
fraction of molecular condensate is
$\sim 10^{-6}$. (b) Results for
$\dot{B}^{-1}=100\text{ $\mu$s/G}$ are typical, and
suggest that the
system undergoes collective adiabatic following from
BEC to dissociated
atom pairs.}
\label{SWEEP}
\end{figure}

{\it Results}-- We begin with the Cornish
\etal~\cite{COR00} experiments
implementing a sweep of the magnetic field across the
Feshbach
resonance, which is of course a version of the age-old
Landau-Zener
problem~\cite{JAV99,KOS00,LZP}. Although a sweep of
the detuning
$\delta$ from above to below threshold at a rate slow
compared to
the condensate coupling $\Omega$ will move the system
adiabatically
from all atoms to all molecules, rogue dissociation
will overtake
coherent atom-molecule conversion when
$\Omega\gsimeq\omega_\rho$~\cite{JAV99,KOS00,JAV02}.
Nevermind that
the JILA experiments sweep from below to above
threshold, for a density
$\rho=1\times 10^{12}\,{\rm cm}^{-3}$ the condensate
coupling is
$\Omega=1.93\times10^5\,\text{s}^{-1}\approx
250\,\omega_\rho$, and so rogue dissociation should
seriously dominate. This is indeed the case
(see Fig.~\ref{SWEEP}). Apparently,
coherent conversion occurs
not between atomic and molecular BEC, but between
atomic BEC and
dissociated atom pairs. Holding this thought, we
conclude that mean-field
theory indicates rogue dissociation as a primary sink of
atoms in the Ref.~\cite{COR00} sweeps across the
Feshbach
resonance.

\begin{figure}
\centering
\epsfig{file=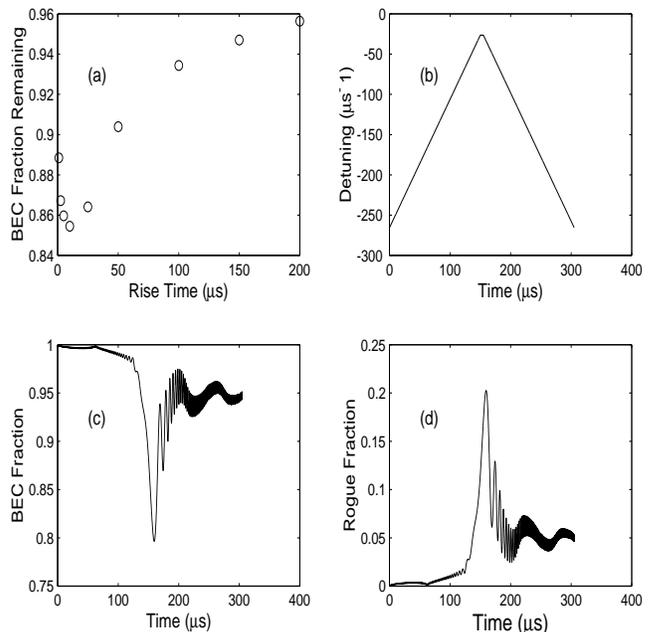,width=8.5cm,height=8.5cm}
\caption {Theory of a magnetic
field pulse applied
to a \rb condensate for
$\rho=1.9\times10^{13}\,\text{cm}^{-3}$ and
$\Omega=8.42\times10^5\,\text{s}^{-1}$. (a)~Remnant
fraction
versus detuning (magnetic field) rise time.
(b-d) Results for a pulse with
$150\,\mu$s rise time indicate adiabatic passage of
BEC atoms to
and from dissociated atom pairs. The minimum in
panel~(a), similar to Ref.~\protect\cite{CLA02}, occurs at the onset of
adiabaticity.}
\label{PULSE}
\end{figure}

Next we consider the experiments of Claussen
\etal~\cite{CLA02}, for which
nontrivial electromagnetic coil technology was
developed to create
trapezoidal magnetic field pulses that bring the
system near-- but
not across-- resonance, hold for a given amount of
time,
and return to the
original field value. Neglecting the burst, these
remnant-focused
experiments revealed a contradiction with the
conventional understanding
of condensate loss: rather than a loss that increased
monotonically with
increasing interaction time, the results
indicated a loss that
increased with {\em decreasing} interaction
time, until very short times were reached.
The present mean-field approach works similarly,
as shown in Fig.~\ref{PULSE}. Our interpretation is
that
adiabaticity is again at play. At very short
pulse durations, increasing interaction time leads to
increasing condensate loss, as expected. In contrast,
as the time dependence of the pulse gets slower, the
system eventually follows the pulse adiabatically,
and returns close to the initial condensate state
when the pulse has passed.

\begin{figure}
\centering
\epsfig{file=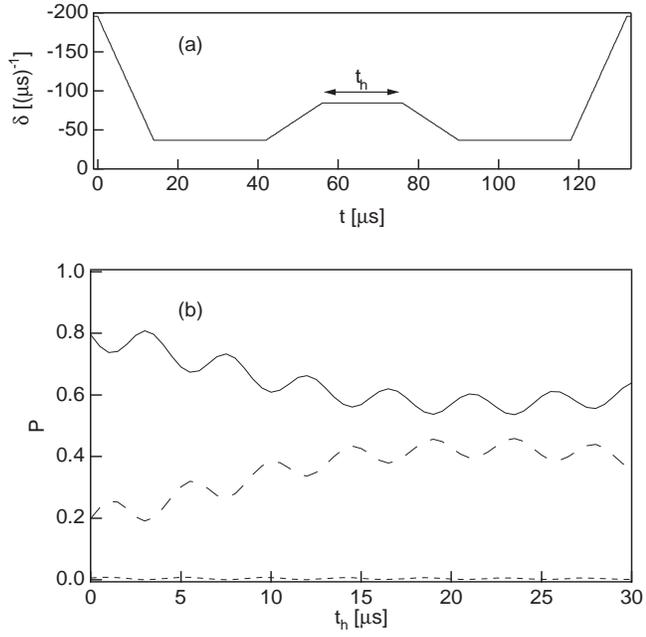,width=8.5cm,height=8.5cm}
\caption{
Simulation of the Ref.~\protect\cite{DON02}
experiments for a
density
$\rho=5.4\times10^{13}\text{cm}^{-3}$
and $\Omega=$ $1.42\times10^6\text{s}^{-1}$. (a) Time
dependence of the
detuning, and (b) the fraction of atoms in the remnant
condensate (solid
line), in noncondensate atoms pairs (dashed line) and
in the molecular
condensate (short-dashed line) after the pulse
sequence
as a function of the
hold time $t_h$ between the two trapezoidal pulses.
The frequency of the
oscillations is compatible with the prediction from
Eq.~\eq{SIGEQ},
identifying these oscillations as Ramsey fringes in
the transition
between the atomic condensate and a molecular
condensate dressed by
noncondensate atom pairs.}
\label{FRINGE}
\end{figure}

Finally, we turn to the experiments performed by
Donley
\etal~\cite{DON02}, in which two trapezoidal pulses
were
applied to a \rb condensate, and the fraction of
remnant and burst
atoms measured for a variable between-pulse time and
magnetic-field amplitude. These
experiments revealed coherent remnant-burst
oscillations with amplitudes
of up to
$\sim 25$\%. As it happens, we have recently predicted
coherent
oscillations between atoms and dissociated atom pairs
in a
rogue-dominated system, although we harbored doubts
regarding any
practical realization~\cite{JAV02}. Casting these
doubts aside, we
consider a time dependent detuning (magnetic field) in
essence lifted
from Fig.~2 of Ref.~\cite{DON02}
[Fig.~\ref{FRINGE}(a)], and determine
the fraction of remnant condensate atoms,
noncondensate atoms, and
molecules at the end of the pulse sequence as a
function of the holding
time between the two pulses [Fig.~\ref{FRINGE}(b)].
Oscillations are seen with the amplitude of about 15\%
between condensate
and noncondensate atoms at the frequency of the
molecular state
corresponding to the magnetic field during the holding
period. The
molecular fraction appears too small to account for
the amplitude of the
oscillations. In fact, what we termed molecular
frequency is the
characteristic frequency of a coherent superposition
of
molecules and
noncondensate atom pair. Here the oscillations,
directly
comparable to Fig.~4(a) in Ref.~\cite{DON02}, are
Ramsey
fringes~\cite{RAM50} in the evolution between an
atomic condensate
and a molecular condensate dressed with noncondensate
atom pairs.

Although our rogue-dissociation ideas provide a neat
qualitative explanation for the three experiments we
have discussed~\cite{ENCON}, in all fairness it must
be noted
that we have fallen short of a full quantitative
agreement. It appears that our model is
missing a so far unidentified additional loss
mechanism for the condensate atoms.

{\it Conclusions}-- We have demonstrated that a
minimal mean-field model is sufficient to
qualitatively
explain a number of puzzling results in
Feshbach-resonant systems. On the
whole, collapsing-condensate physics is therefore
understood as a matter
of rogue dissociation, which leads to strong losses in
the threshold
neighborhood, decreased remnant fraction for
decreasing
interaction time---until very short times are reached,
and coherent remnant-burst oscillations. Furthermore,
although atom-molecule coherence has no doubt been
achieved~\cite{DON02}, the
amplitude of the remnant-burst
oscillations need not be
indicative of the number of
condensate molecules present. Ironically, the
strength of the
Feshbach resonance has led to a regime dominated by
rogue
dissociation, which tends to counteract the production
of a
molecular condensate.

{\it Acknowledgements}-- We thank Neil
Claussen and Eddy
Timmermans for helpful discussions, and the Academy of
Finland (MM and
KAS, projects 43336 and 50314), NSF (JJ, Grants
PHY-9801888 and
PHY-0097974), and NASA (JJ, Grant
NAG8-1428) for support.

\vspace{-0.25cm}

\begin{references}
\bibitem{ENCON}
   If an explanation in terms of below-threshold
molecular dissociation
seems a bit odd, consider that energy need not be
conserved for transient
processes where a time dependence is externally
imposed on the system.
\end{references}
\end{document}